\documentclass[
 superscriptaddress,
 amsmath,amssymb,
 aps,
 twocolumn,
prc,
]{revtex4-1}

\usepackage{graphicx}
\usepackage{dcolumn}
\usepackage{bm}
\usepackage[hyperindex,breaklinks]{hyperref}
\hypersetup{
    colorlinks,
    citecolor=blue,
    filecolor=black,
    linkcolor=blue,
    urlcolor=blue
}
\usepackage{subfigure}
\usepackage[mathlines]{lineno}
\usepackage{breqn}
\usepackage{amsmath}

\def\BBz{0$\nu\beta\beta$}

\def\mj{M{\sc ajo\-ra\-na}}
\def\dem{D{\sc e\-mon\-strat\-or}}

\def\QBB{Q$_{\beta\beta}$}

\def\ge{$^{76}$Ge}

\begin{document}

\preprint{APS/123-QED}

\title{Multi-site event discrimination for the {\sc Majorana Demonstrator}}


\newcommand{\blhill}{Department of Physics, Black Hills State University, Spearfish, SD, USA}
\newcommand{\ITEP}{National Research Center ``Kurchatov Institute'' Institute for Theoretical and Experimental Physics, Moscow, Russia}
\newcommand{\JINR}{Joint Institute for Nuclear Research, Dubna, Russia}
\newcommand{\lbnl}{Nuclear Science Division, Lawrence Berkeley National Laboratory, Berkeley, CA, USA}
\newcommand{\lanl}{Los Alamos National Laboratory, Los Alamos, NM, USA}
\newcommand{\queens}{Department of Physics, Engineering Physics and Astronomy, Queen's University, Kingston, ON, Canada}
\newcommand{\uw}{Center for Experimental Nuclear Physics and Astrophysics, 
and Department of Physics, University of Washington, Seattle, WA, USA}
\newcommand{\unc}{Department of Physics and Astronomy, University of North Carolina, Chapel Hill, NC, USA}
\newcommand{\duke}{Department of Physics, Duke University, Durham, NC, USA}
\newcommand{\ncsu}{Department of Physics, North Carolina State University, Raleigh, NC, USA}	
\newcommand{\ornl}{Oak Ridge National Laboratory, Oak Ridge, TN, USA}
\newcommand{\ou}{Research Center for Nuclear Physics, Osaka University, Ibaraki, Osaka, Japan}
\newcommand{\pnnl}{Pacific Northwest National Laboratory, Richland, WA, USA}
\newcommand{\princeton}{Department of Physics, Princeton University, Princeton, NJ, USA}
\newcommand{\ttu}{Tennessee Tech University, Cookeville, TN, USA}
\newcommand{\sdsmt}{South Dakota School of Mines and Technology, Rapid City, SD, USA}
\newcommand{\usc}{Department of Physics and Astronomy, University of South Carolina, Columbia, SC, USA}
\newcommand{\usd}{Department of Physics, University of South Dakota, Vermillion, SD, USA} 
\newcommand{\ut}{Department of Physics and Astronomy, University of Tennessee, Knoxville, TN, USA}
\newcommand{\tunl}{Triangle Universities Nuclear Laboratory, Durham, NC, USA}
\newcommand{\mpi}{Max-Planck-Institut f\"{u}r Physik, M\"{u}nchen, Germany}
\newcommand{\tum}{Physik Department, Technische Universit\"{a}t, M\"{u}nchen, Germany}
\newcommand{\MIT}{Department of Physics, Massachusetts Institute of Technology, Cambridge, MA, USA} 

\affiliation{\uw}
\affiliation{\pnnl}
\affiliation{\usc}
\affiliation{\ornl}
\affiliation{\ITEP}
\affiliation{\usd}
\affiliation{\mpi}
\affiliation{\JINR}
\affiliation{\duke}
\affiliation{\tunl}
\affiliation{\uw}
\affiliation{\unc}
\affiliation{\lbnl}
\affiliation{\sdsmt}
\affiliation{\lanl}
\affiliation{\ut}
\affiliation{\ou}
\affiliation{\princeton}
\affiliation{\ncsu}
\affiliation{\MIT}
\affiliation{\blhill}
\affiliation{\ttu}
\affiliation{\queens} 
\affiliation{\tum}

\author{S.I.~Alvis}\affiliation{\uw}	
\author{I.J.~Arnquist}\affiliation{\pnnl} 
\author{F.T.~Avignone~III}\affiliation{\usc}\affiliation{\ornl}
\author{A.S.~Barabash}\affiliation{\ITEP}
\author{C.J.~Barton}\affiliation{\usd}	
\author{F.E.~Bertrand}\affiliation{\ornl}
\author{B.~Bos}\affiliation{\sdsmt} 
\author{M.~Buuck}\affiliation{\uw}  
\author{T.S.~Caldwell}\affiliation{\unc}\affiliation{\tunl}	
\author{Y-D.~Chan}\affiliation{\lbnl}
\author{C.D.~Christofferson}\affiliation{\sdsmt} 
\author{P.-H.~Chu}\affiliation{\lanl} 
\author{C. Cuesta}	\email{clara.cuesta@ciemat.es}\altaffiliation{Present address: Centro de Investigaciones Energ\'{e}ticas, Medioambientales y Tecnol\'{o}gicas, CIEMAT 28040, Madrid, Spain}\affiliation{\uw}
\author{J.A.~Detwiler}\affiliation{\uw}	
\author{H.~Ejiri}\affiliation{\ou}
\author{S.R.~Elliott}\affiliation{\lanl}
\author{T.~Gilliss}\affiliation{\unc}\affiliation{\tunl}  
\author{G.K.~Giovanetti}\affiliation{\princeton}  
\author{M.P.~Green}\affiliation{\ncsu}\affiliation{\tunl}\affiliation{\ornl}   
\author{J.~Gruszko}\affiliation{\MIT} 
\author{I.S.~Guinn}\affiliation{\uw}		
\author{V.E.~Guiseppe}\affiliation{\usc}	
\author{C.R.~Haufe}\affiliation{\unc}\affiliation{\tunl}	
\author{R.J.~Hegedus}\affiliation{\unc}\affiliation{\tunl} 
\author{L.~Hehn}\affiliation{\lbnl}	
\author{R.~Henning}\affiliation{\unc}\affiliation{\tunl}
\author{D.~Hervas~Aguilar}\affiliation{\unc}\affiliation{\tunl} 
\author{E.W.~Hoppe}\affiliation{\pnnl}
\author{M.A.~Howe}\affiliation{\unc}\affiliation{\tunl}
\author{K.J.~Keeter}\affiliation{\blhill}
\author{M.F.~Kidd}\affiliation{\ttu}	
\author{S.I.~Konovalov}\affiliation{\ITEP}
\author{R.T.~Kouzes}\affiliation{\pnnl}
\author{A.M.~Lopez}\affiliation{\ut}	
\author{R.D.~Martin}\affiliation{\queens}	
\author{R.~Massarczyk}\affiliation{\lanl}		
\author{S.J.~Meijer}\affiliation{\unc}\affiliation{\tunl}	
\author{S.~Mertens}\affiliation{\mpi}\affiliation{\tum}		
\author{J.~Myslik}\affiliation{\lbnl}		
\author{G.~Othman}\affiliation{\unc}\affiliation{\tunl} 
\author{W.~Pettus}\affiliation{\uw}	
\author{A.~Piliounis}\affiliation{\queens} 
\author{A.W.P.~Poon}\affiliation{\lbnl}
\author{D.C.~Radford}\affiliation{\ornl}
\author{J.~Rager}\affiliation{\unc}\affiliation{\tunl}	
\author{A.L.~Reine}\affiliation{\unc}\affiliation{\tunl}	
\author{K.~Rielage}\affiliation{\lanl}
\author{N.W.~Ruof}\affiliation{\uw}	
\author{B.~Shanks}\affiliation{\ornl}	
\author{M.~Shirchenko}\affiliation{\JINR}
\author{D.~Tedeschi}\affiliation{\usc}		
\author{R.L.~Varner}\affiliation{\ornl}  
\author{S.~Vasilyev}\affiliation{\JINR}	
\author{Vasundhara}\affiliation{\queens} 
\author{B.R.~White}\affiliation{\lanl}	
\author{J.F.~Wilkerson}\affiliation{\unc}\affiliation{\tunl}\affiliation{\ornl}    
\author{C.~Wiseman}\affiliation{\uw}		
\author{W.~Xu}\affiliation{\usd} 
\author{E.~Yakushev}\affiliation{\JINR}
\author{C.-H.~Yu}\affiliation{\ornl}
\author{V.~Yumatov}\affiliation{\ITEP}
\author{I.~Zhitnikov}\affiliation{\JINR} 
\author{B.X.~Zhu}\affiliation{\lanl} 
			
\collaboration{{\sc{Majorana}} Collaboration}
\noaffiliation

\date{\today}

\begin{abstract}

The  \mj\ \dem\ is searching for neutrinoless double-beta decay (\BBz) in \ge\ using arrays of point-contact germanium detectors operating at the Sanford Underground Research Facility. Background results in the \BBz\ region of interest from data taken during construction, commissioning, and the start of full operations have been recently published. A pulse shape analysis cut applied to achieve this result, named $AvsE$, is described in this paper. This cut is developed to remove events whose waveforms are typical of multi-site energy deposits while retaining (90\,$\pm$3.5)\% of single-site events. This pulse shape discrimination is based on the relationship between the maximum current and energy, and tuned using $^{228}$Th calibration source data. The efficiency uncertainty accounts for variation across detectors, energy, and time, as well as for the position distribution difference between calibration and \BBz\ events, established using simulations.

\begin{description}
\item[PACS numbers]
\verb+23.40-s, 23.40.Bw, 14.60.Pq+
\end{description}
\end{abstract}

\pacs{Valid PACS appear here}
\maketitle


\section{Introduction}
\label{sec1}

The \mj\ Collaboration is operating an array of high purity Ge (HPGe) detectors to search for neutrinoless double-beta decay (\BBz) in \ge~\cite{mjd,mjd0nbb}. The \mj\ 
\dem\ is comprised of HPGe detectors with a total mass of 44.1\,kg, 29.7\,kg of which is enriched to 88\% in \ge\ and the remaining 14.4\,kg is natural Ge (7.8\% \ge). P-type point contact (PPC) detectors~\cite{ppc,ppc2} were chosen after extensive R\&D by the collaboration for their powerful background rejection capabilities. The detectors are operated near liquid nitrogen temperature (77\,K) in independent vacuum cryostats, named Modules~1 and~2. A low-mass front-end (LMFE) electronic board is situated adjacent to the each detector inside the vacuum cryostat to minimize the readout noise~\cite{mjd,osti_1202433}.  A 2.15\,m signal cable connects the LMFE with the preamplifiers located outside the cryostat.  The signals are then digitized at 100\,MHz by a 14-bit ADC. The modules are operated in a low-background passive shield that is surrounded by a 4$\pi$ active muon veto. To mitigate the effect of cosmic rays and prevent cosmogenic activation of detectors and materials, the experiment is operating at a depth of 4850\,ft (4260\,m.w.e.\,\,overburden) at the Sanford Underground Research Facility in Lead, South Dakota, USA~\cite{surf}.

We presented results from data taken over June 2015 - April 2018, a 26\,kg\,yr exposure, including construction, commissioning, and stable full operation.  An unprecedented energy resolution of 2.5\,keV FWHM at the \BBz\ Q-value  (\QBB\,= 2039\,keV for \ge) was achieved. Also, a very low background was reached with a single candidate event in the optimal region of interest (ROI) resulting in a lower limit on the half-life of 2.7\,$\times$\,10$^{25}$\,yr (90\% CL)~\cite{mjd0nbb,mjd0nbb18}. In our experimental configuration with the lowest background, the background is 11.9$\pm$2.0\,counts/(FWHM\,t\,yr). In order to achieve this low background, multi-site background events are rejected with the method and efficiency described in this paper. 

The data presented in the \BBz\ results are subdivided into data-sets, referred to as DS0 through DS6, distinguished by significant experimental configuration changes. DS0 was a set of commissioning runs of Module~1. DS1 had the inner 2-inch electroformed copper shield installed. DS2 was devoted to test multisampling of the digitized waveforms, providing extended signal capture following an event for improved alpha background rejection. DS3 and DS4 consist of data taken from Module~1 and Module~2, respectively, with separate DAQ systems.
DS5 consists of three sub-ranges corresponding to minor configuration changes.  DS5a was marked by combined data taking with both modules after the DAQ systems were merged. DS5b corresponds to data taken after the detector was fully enclosed within the layer of poly shielding, allowing the establishment of a robust grounding scheme that reduced the electronic noise. DS5c implemented blindness and was excluded from the first result analysis. Finally, in DS6 multisampling is in place.

\section{Multi-site event discrimination in PPC detectors}
\label{sec2}

\begin {figure}[ht]
\includegraphics[width=0.45\textwidth]{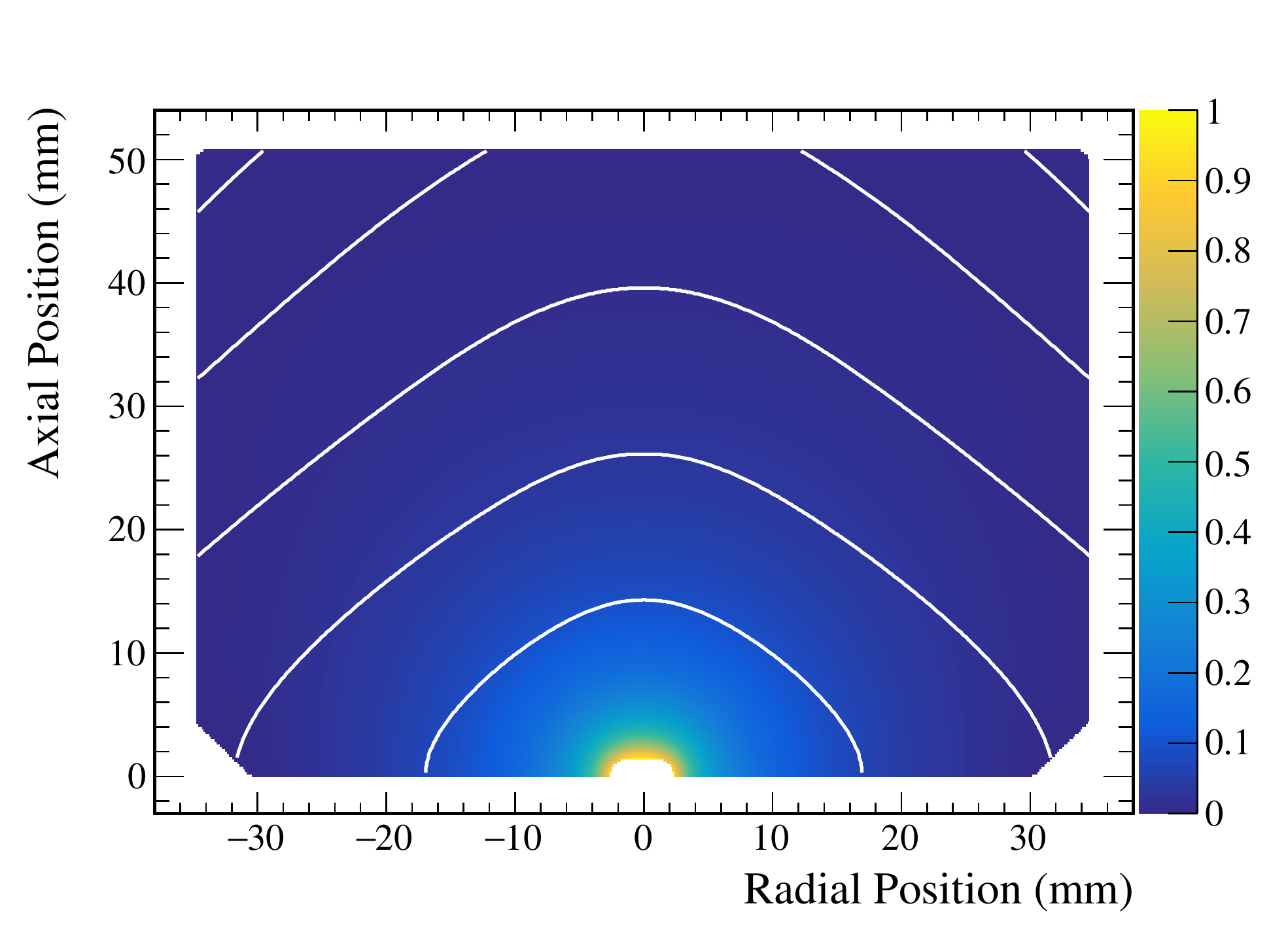}
\centering \caption{\it  Weighting potential for the point contact (bottom center) in a PPC detector. White lines are isochrones of equal drift time for holes to reach the point contact spaced by 200\,ns.}
\label{fig:wp}
\end {figure}

\begin {figure}[ht]
\includegraphics[width=0.45\textwidth]{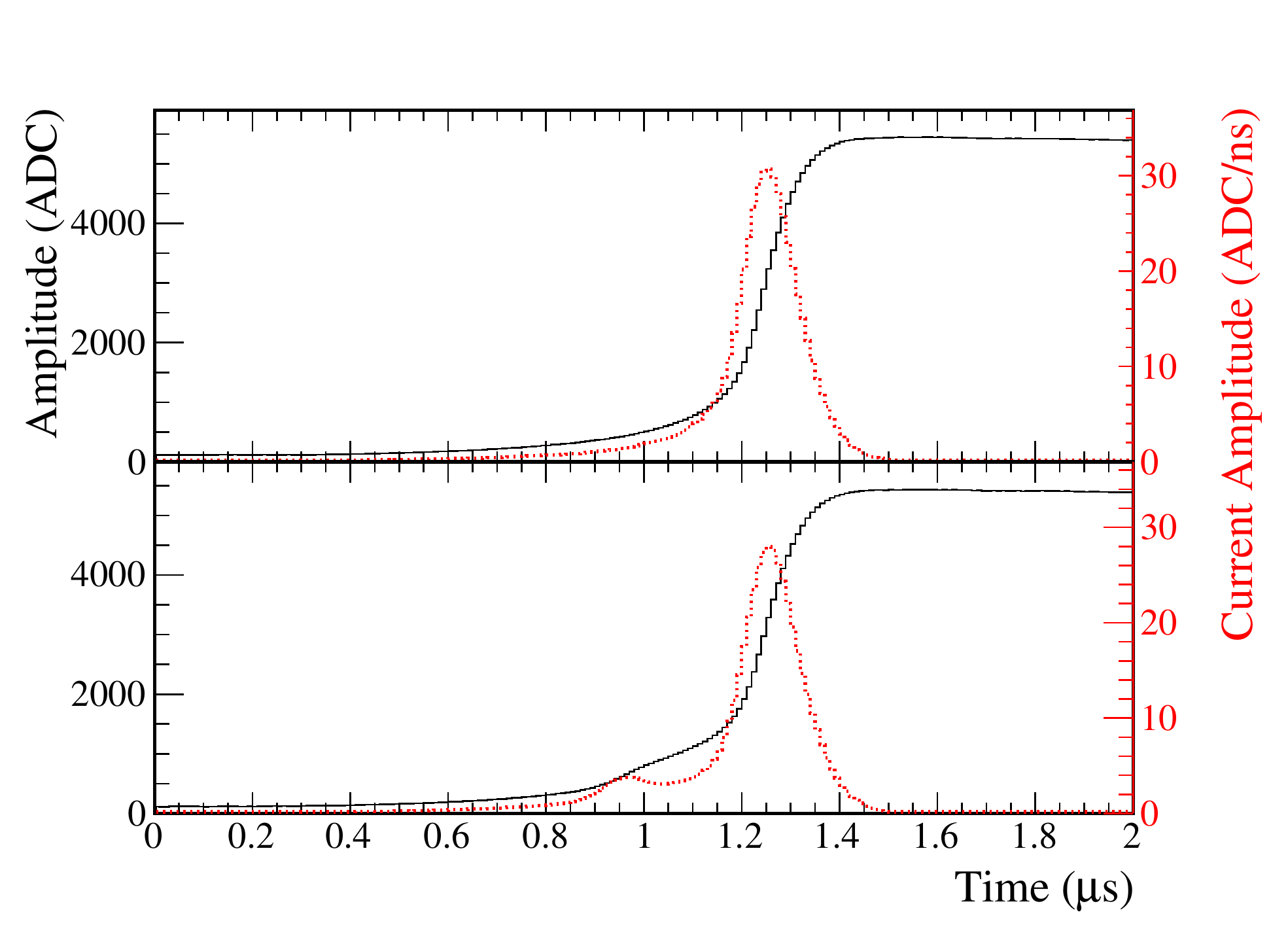}
\centering \caption{\it  Charge (black, solid) and current (red, dashed) signals formed by SSE (top) and MSE (bottom) in a PPC detector. Both events have near \QBB\ energy and are from experimental data.}
\label{fig:pulse}
\end {figure}

The experimental sensitivity is improved by pulse shape analysis (PSA) of the detector signals to reject background events. In particular, the \BBz\ event topology consists of the two electrons carrying the entire decay energy.  This results in a monoenergetic peak at the \QBB, with all the energy being deposited within $\sim$1\,mm in a single-site energy deposit.  Therefore, single-site events (SSE) must be retained, but multi-site events (MSE) characteristic of gamma backgrounds should be rejected. The point contact detector technology was chosen for the strong weighting potential in the vicinity of the point contact readout and the relatively low weighting potential elsewhere throughout the detector, see Fig.~\ref{fig:wp}. This forces the majority of the charge to be collected only at the very end of the trajectory of the charge drift within the detector resulting in a signal that has a risetime that is much shorter than the drift time of charge through the detector. If charge is deposited at multiple locations within the crystal, the drift times may differ up to $\sim$1\,$\mu$s and the individual charge collections can be resolved. This leads to a signal with a current pulse that is degraded in amplitude with respect to the current pulse relative to that of a SSE of the same energy. Examples of current and charge pulses for SSE and MSE are shown in Fig.~\ref{fig:pulse}. By comparing the maximum amplitude of the current pulse ($A$) with the energy ($E$) we can reject events that have a spread-out current pulse and are likely multi-site as indicated by low values of $A$ relative to $E$~\cite{AEgerda}. 

As a reference population of SSE, we use the Double Escape Peak (DEP) of the 2614-keV $^{208}$Tl gamma ray. This peak is generated by the creation of an electron positron pair during the photon interaction with a nucleus of the detector. The photons from the positron annihilation both escape the detector leaving an energy deposit 
1592\,keV, two electron masses less than the incident gamma ray energy. This physics requires these events to have single-site structure similar to that expected of 0$\nu\beta\beta$. Monte Carlo simulations including X-ray excitations and bremsstrahlung predict the 0$\nu\beta\beta$ signal events to be 90\% single-site. Defining a cut to leave this fraction of events in the DEP yields the near-optimal rejection efficiencies for the single escape peak (SEP) at 2103~keV (mostly MSE) and the Compton continuum in the ROI. A cut to remove high values of $A$ relative to $E$, which is functionally a fiducial volume cut targeting around the point contact, is not applied to the data as it performs a largely redundant function to the delayed charge recovery (DCR) cut~\cite{dcr}, but with lower signal efficiency.

\section{The $AvsE$ parameter}
\label{sec3}

In order to create an energy-independent parameter, the $AvsE$ parameter is calculated considering the energy dependence of $A$. Event energies are reconstructed from the pulse amplitudes, using a trapezoidal filter algorithm whose parameters are tuned to minimize calibration source gamma line widths ~\cite{knollpaper}. The current estimator is an algorithm that performs a linear fit to a small range of the waveform. Since the pre-amplifiers used to record the waveforms in our detectors are charge sensitive, it is critical to this analysis to have an accurate estimate of the current from the digitized charge waveform. Three differentiation time constants (50\,ns, 100\,ns and 200\,ns) were considered. Very similar performance was observed for each parameter  and 100\,ns was selected as the time constant for the $A$ estimator.

The energy dependence of $A$ is observed to be second-order polynomial that is mostly linear with a small quadratic component.  $AvsE$ is thus defined,

\begin{equation}
AvsE \equiv -1\cdot(A\cdot E/E_{unc}-p0-p1\cdot E -p2\cdot E^{2})/j
\end{equation}
\label{eqn:avse}

\noindent where $p0$, $p1$, and $p2$ are the energy dependence parameters, $E$ and $E_{unc}$ are calibrated and uncalibrated energy, and $j$ sets the cut value. Events with $AvsE>-1$ have SSE character and are accepted.

As $A$ is uncalibrated, it is multiplied by $E/E_{unc}$ to account for gain shifts and to be able to compare it to $E$. $A$ is based on a slope across 10 waveform samples, so its distribution naturally has larger width ($\sim$1\%) than energy ($\sim$0.1\%). A non-linear $A$ dependence with $E$ is expected to arise from the spatial energy deposition (higher energy betas having a larger range and thus larger initial ionization distribution), space charge effects (repulsion of charges broaden the distribution during drift to the electrodes), and response of the electronics.  All these effects work to introduce a negative quadratic term, which is equivalent to the negative linear term in the $A/E$ study of~\cite{AEgerda}.

\begin {figure}[ht]
\includegraphics[width=0.45\textwidth]{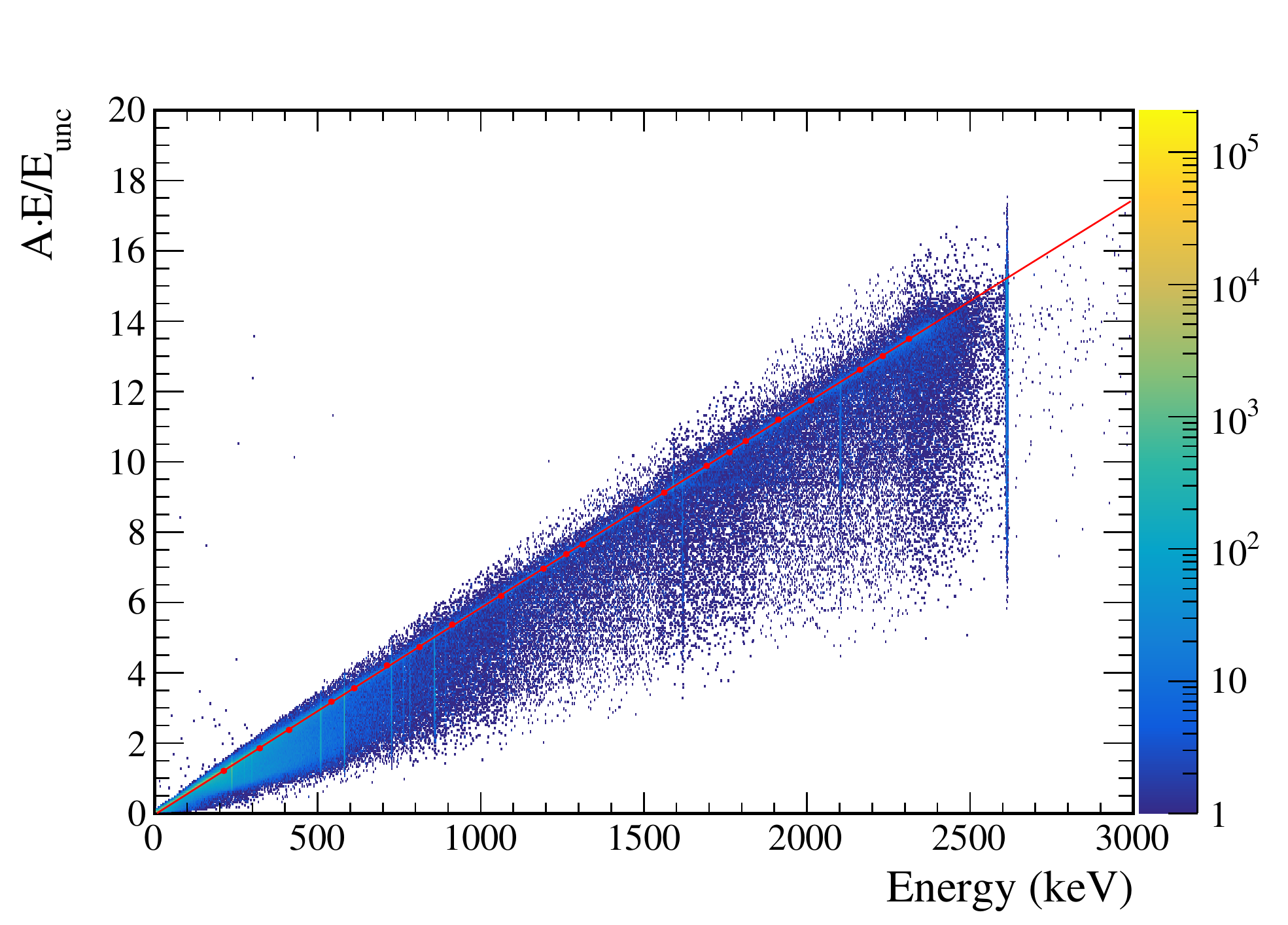}
\centering \caption{\it  The distribution of $A$ vs $E$ for a detector. The red dots are the mode of $A\cdot E/E_{unc}$ at the evaluated energies and the red line is the quadratic polynomial fit 
$A\cdot E/E_{unc} = -3.904\cdot10^{-2} + 5.908\cdot10^{-3} \cdot E-2.665\cdot 10^{-8} \cdot E^{2}$.}
\label{fig:avse}
\end {figure}

\begin {figure}[ht]
\subfigure[]{\includegraphics[width=0.45\textwidth]{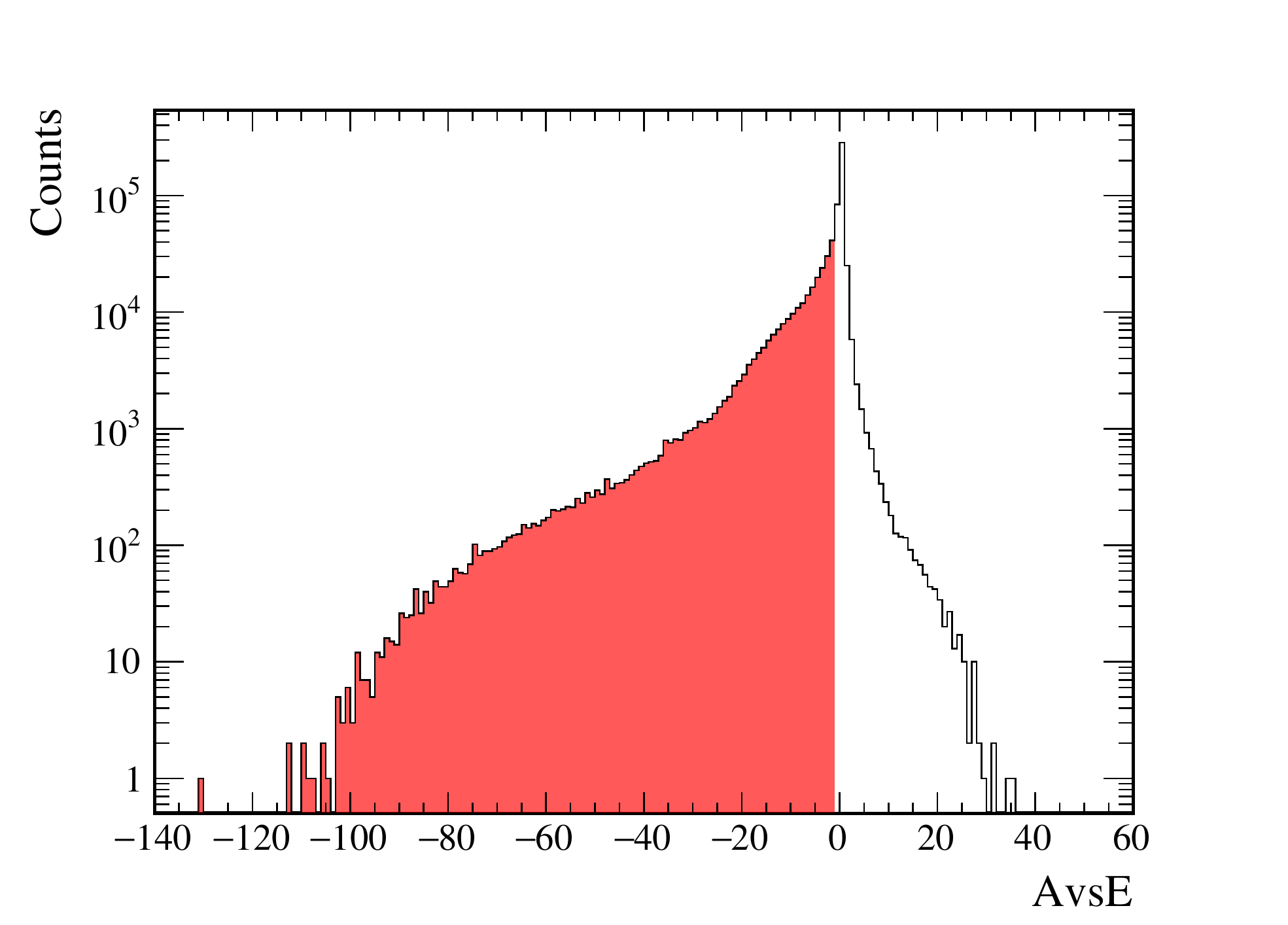}\label{fig:avse2a}}
\subfigure[]{\includegraphics[width=0.45\textwidth]{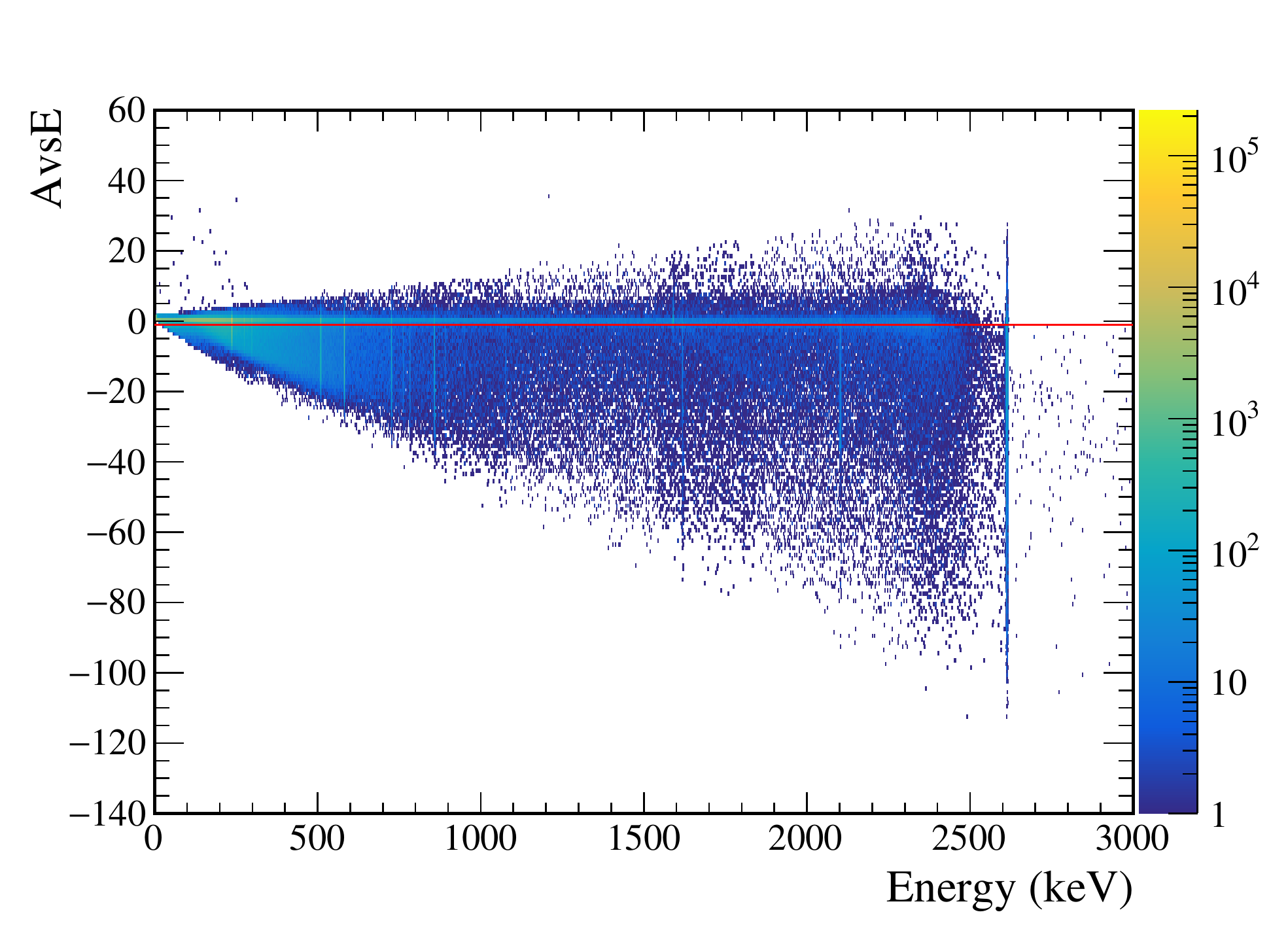}\label{fig:avse2b}}
\centering \caption{\it (a) The corrected A value, also known as $AvsE$, for E$>$100\,keV with the excluded MSE shaded in red.  (b) $AvsE$ vs Energy with excluded MSE below the red line.}
\label{fig:avse2}
\end {figure}

To calculate the cut parameters, the following methodology is applied. First, 22 Compton-continuum regions from 200-2300\,keV, each 25\,keV wide, are considered. For each region, the mode of the $A \cdot E/E_{unc}$ distribution is obtained.  A quadratic fit to these 22 points is applied to get $p0$, $p1$, and $p2$, see Fig.~\ref{fig:avse}. Finally, the $j$ parameter is varied until 90\% of background-subtracted DEP events pass the cut. The corrected $A$ value, also known as $AvsE$, is shown in Fig.~\ref{fig:avse2a} and as a function of the energy in Fig.~\ref{fig:avse2b}. 

\begin {figure}[ht]
\includegraphics[width=0.45\textwidth]{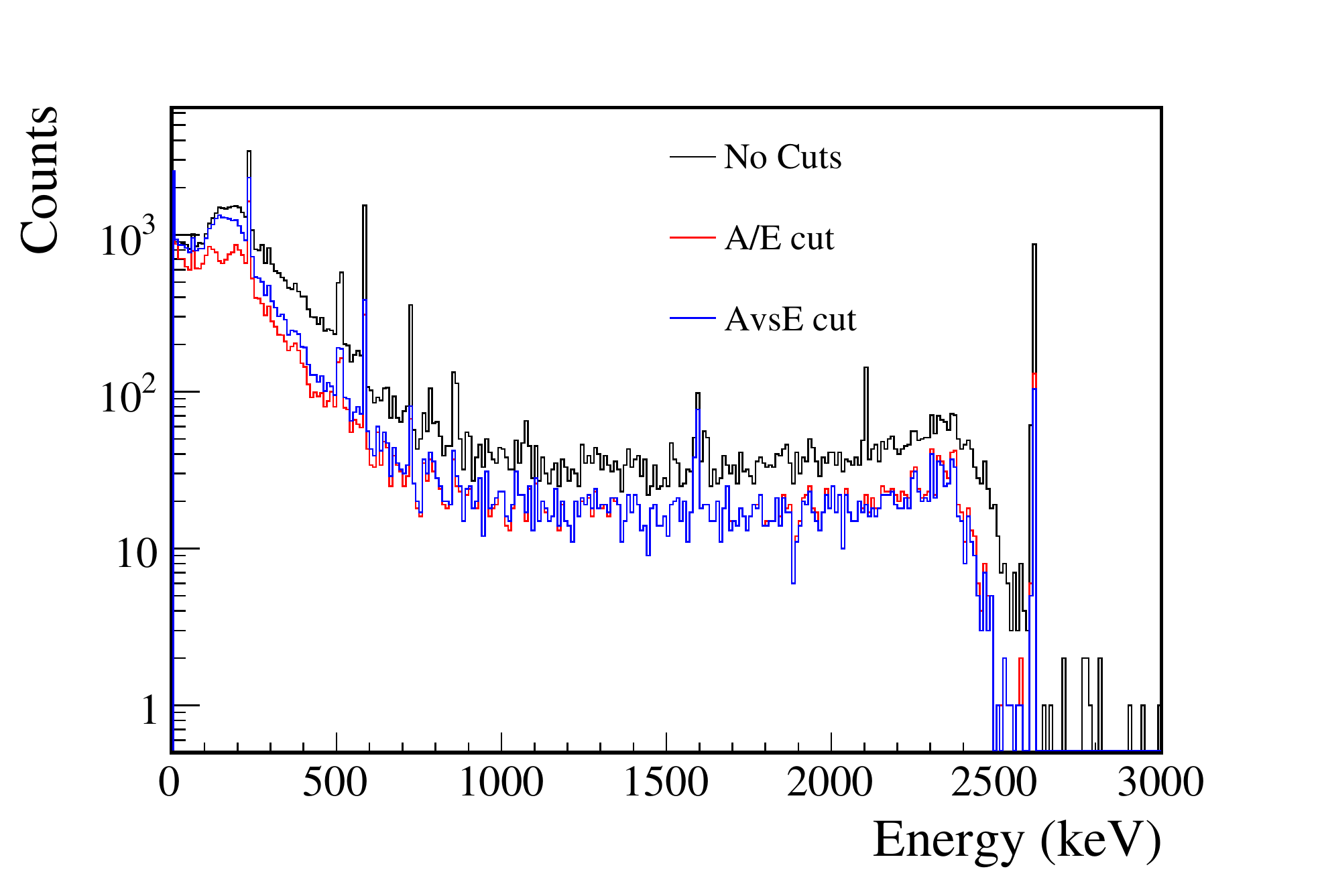}
\centering \caption{\it  Comparison of the $A/E$ and the $AvsE$ cuts applied to $^{228}$Th calibration data.}
\label{fig:aovere}
\end {figure}

The possibility of using the peak amplitude to total energy, $A/E$, as the cut parameter (instead of $AvsE$)~\cite{AEgerda} was also explored. However, the width of $A/E$ increases significantly at lower energies, reducing the efficiency for SSE~\cite{AEgerda}. A 1\,MeV energy cut had to be applied to achieve a constant $A/E$ cut performance. Although this threshold is far below the 0$\nu\beta\beta$ region of interest, other spectral analyses require a lower threshold for the multi-site event cut. The $AvsE$ cut has demonstrated performance in the \mj\ \dem\ down to 100\,keV, below which noise events become the dominant background requiring other cuts~\cite{mjdle}. A comparison of both cuts is shown in Fig.~\ref{fig:aovere}.

\section{Efficiency determination using $^{228}$Th calibration data}
\label{sec4}

We calibrate the detectors with a $^{228}$Th line source~\cite{calibration}. At least one long ($\sim$12\,h) $^{228}$Th calibration is taken during each data-set to ensure enough statistics ($\mathcal{O}(1000)$~DEP events/detector) to individually calibrate the $AvsE$ parameters for each detector. More frequent short $\sim$1\,hr calibrations are used to monitor time stability, but have insufficient statistics to individually calibrate the $AvsE$ parameters. For each data-set, the $AvsE$ acceptance value is set so that the survival efficiency of the DEP is 90\%. Then, the survival efficiencies are calculated for the SEP and for the 100-keV region surrounding \QBB\ where most events are Compton scattered recoil electrons. Multiple long calibration runs were taken in DS0, DS1, and DS6, and the cut is recalculated for each. The same long calibration is used for DS5a, DS5b, and DS5c, as no different results are expected. 

To determine the efficiency, we compute the total number of events in the DEP (SEP) window ($N$) and in the background window ($B$) and the number of events passing ($N_c$, $B_c$) the $AvsE$ cut ($AvsE >$ -1). The signal energy windows for the DEP (SEP) are 1590-1595\,keV (2101-2106\,keV) giving $N$ and $N_c$, and the background energy windows are 1570-1580\,keV and 1600-1610\,keV (2080-2090\, keV and 2115-2125\,kV) giving $B$ and $B_c$. We compute the efficiency via background-subtraction:

\begin{equation}
\label{eqn:eff}
\epsilon = \frac{N_c - \tau B_c}{N - \tau B}
\end{equation}

\noindent where $\tau$ is the energy width ratio between the signal and background windows. The uncertainty $\sigma_\epsilon$ is computed by standard error propagation, accounting for the covariance between $N$ and $N_c$, and $B$ and $B_c$:

\begin{equation}
\label{eqn:stat}
\begin{split}
\left(\frac{\sigma_\epsilon}{\epsilon} \right)^2 = \frac{N + \tau^2
B}{(N - \tau B)^2} + \frac{N_c + \tau^2 B_c}{(N_c - \tau B_c)^2} \\  - 2
\frac{N_c + \tau^2 B_c}{(N - \tau B)(N_c - \tau B_c)}
\end{split}
\end{equation}

\begin {figure*}[ht]
\subfigure[]{\includegraphics[width=0.95\textwidth]{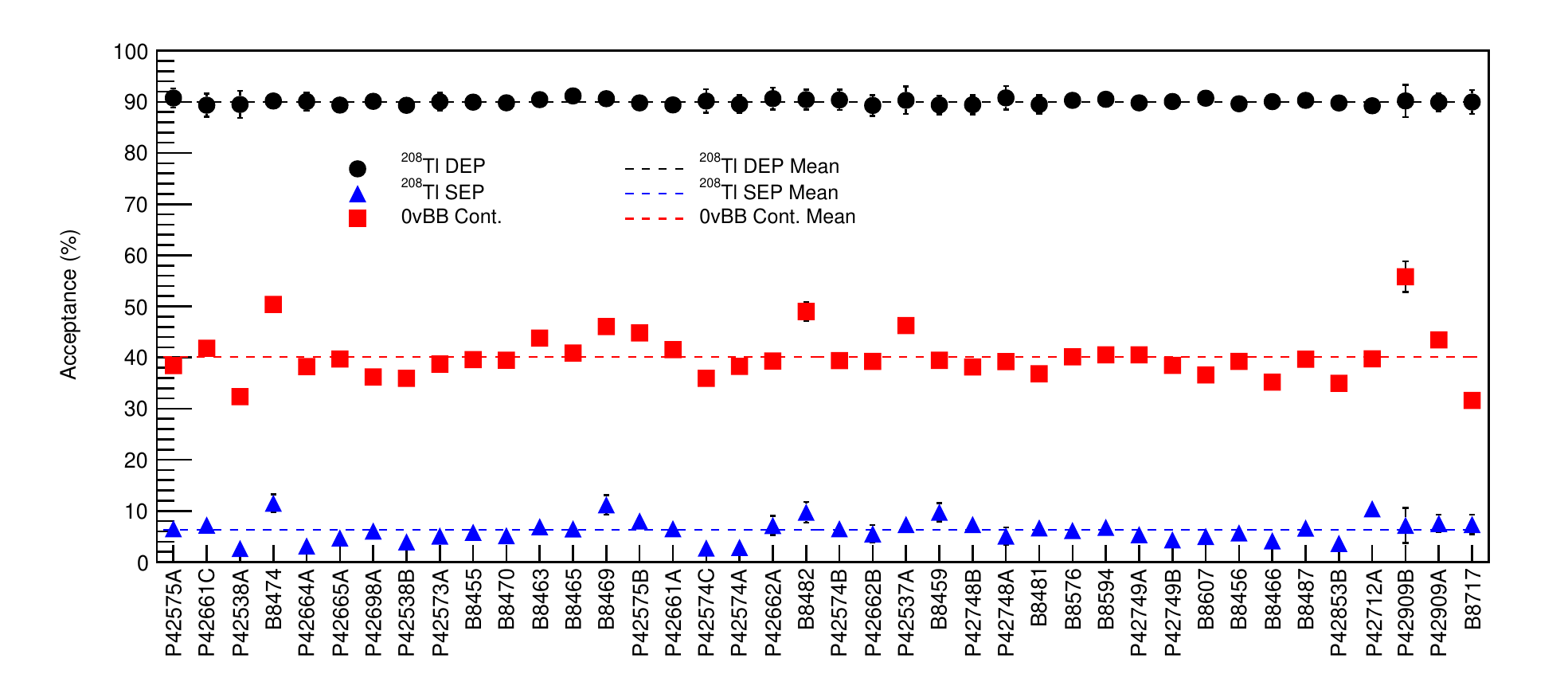}}
\centering \caption{\it $AvsE$ PSA performance of the operating detectors in DS5. The single-site $^{208}$Tl DEP events are fixed to 90\% (black), the multi-site SEP events (blue) are reduced to 6\% by the cut.}
\label{fig:eff}
\end {figure*}

For the ROI where no background-subtraction is relevant, the efficiency is calculated as the ratio of the integral of the 1989 - 2089\,keV energy region after and before the $AvsE$ cut. The percentage of accepted events by detector is shown in Fig.~\ref{fig:eff} for DS5 where both modules were first operative. The average survival efficiency for all data-sets are shown in Table~\ref{tab:mean}. The small deviation in the DEP survival efficiency from the 90\% prescription is mainly due to the statistical uncertainty.

\begin{table}[ht]
\begin{center}
\begin{tabular}{lccc}
\hline\noalign{\smallskip}
Data-set	& DEP (\%)								& SEP (\%)								& ROI (\%)							\\
DS0		& 90.09\,$\pm$\,0.52	& 5.30\,$\pm$\,0.38 	& 38.69\,$\pm$\,0.33\\
DS1		& 90.14\,$\pm$\,0.33 & 5.51\,$\pm$\,0.23	& 39.31\,$\pm$\,0.21\\
DS2		& 90.34\,$\pm$ 0.75 & 6.52\,$\pm$\,0.56 & 42.41\,$\pm$\,0.50\\
DS3		& 89.99\,$\pm$\,0.25 & 5.63\,$\pm$\,0.18	& 39.04\,$\pm$\,0.16\\
DS4		& 89.87\,$\pm$\,0.30 & 7.67\,$\pm$\,0.28	& 41.65\,$\pm$\,0.22\\
DS5 	    & 90.00\,$\pm$\,0.29 & 6.24\,$\pm$\,0.23	& 40.26\,$\pm$\,0.15\\
DS6   	& 90.14\,$\pm$\,0.11  & 6.12\,$\pm$\,0.09 	& 40.21\,$\pm$\,0.06\\
\noalign{\smallskip}\hline
\end{tabular}
\caption{\it Average survival efficiencies of events in the DEP (SSE), SEP (MSE), and ROI (Compton continuum) when subjected to the recommended $AvsE$ cut based on $^{228}$Th calibration. Only the statistical uncertainty is shown.}
\label{tab:mean}
\end{center}
\end{table}

\section{Efficiency Uncertainty}
\label{sec5}

A careful study has been carried out in order to determine the uncertainty associated with the efficiency values calculated in Section~\ref{sec4}. The uncertainty is the quadratic sum of the following components: statistical uncertainty of the DEP survival fraction ($stat$), uncertainty due to $AvsE$ energy dependence ($roi$), uncertainty due to the residual differences between calibration and physics data (2$\nu\beta\beta$), uncertainty due to the difference between $0\nu\beta\beta$ and DEP events, and uncertainty due to time stability ($stab$). The different component contributions are summarized in Table~\ref{tab:sys} and detailed in the following subsections.

\begin{table*}[ht]
\begin{center}
\begin{tabular}{lc}
\hline\noalign{\smallskip}
Data-set	& DEP efficiency and uncertainty\\
\hline\noalign{\smallskip}
DS0 				& 0.9009 $\pm$ 0.0052($stat$) $^{+0.0051}_{-0.0148}$($roi$) $^{+0.0026}_{-0.0046}$($2\nu\beta\beta$) $\pm$ 0.029($0\nu\beta\beta$) $\pm$ 0.0117($stab$) \\
DS1 				& 0.9014 $\pm$ 0.0033($stat$) $^{+0.0033}_{-0.0185}$($roi$) $^{+0.0025}_{-0.0029}$($2\nu\beta\beta$) $\pm$ 0.029($0\nu\beta\beta$) $\pm$ 0.0210($stab$)\\
DS2 				& 0.9034	$\pm$ 0.0075($stat$) $^{+0.0044}_{-0.0148}$($roi$) $^{+0.0068}_{-0.0040}$($2\nu\beta\beta$) $\pm$ 0.029($0\nu\beta\beta$) $\pm$ 0.0187($stab$)\\
DS3 				& 0.8999	$\pm$ 0.0025($stat$) $^{+0.0034}_{-0.0086}$($roi$) $^{+0.0010}_{-0.0012}$($2\nu\beta\beta$) $\pm$ 0.029($0\nu\beta\beta$) $\pm$ 0.0079($stab$)\\
DS4 				& 0.8997	$\pm$ 0.0030($stat$) $^{+0.0039}_{-0.0155}$($roi$) $^{+0.0111}_{-0.0138}$($2\nu\beta\beta$) $\pm$ 0.029($0\nu\beta\beta$) $\pm$ 0.0081($stab$)\\
DS5a       		& 0.9000	$\pm$ 0.0029($stat$) $^{+0.0030}_{-0.0113}$($roi$) $^{+0.0039}_{-0.0047}$($2\nu\beta\beta$) $\pm$ 0.029($0\nu\beta\beta$) $\pm$ 0.0177($stab$)\\
DS5b       		& 0.9000	$\pm$ 0.0029($stat$) $^{+0.0030}_{-0.0113}$($roi$) $^{+0.0039}_{-0.0047}$($2\nu\beta\beta$) $\pm$ 0.029($0\nu\beta\beta$) $\pm$ 0.0118($stab$)\\
DS5c       		& 0.9000	$\pm$ 0.0029($stat$) $^{+0.0030}_{-0.0113}$($roi$) $^{+0.0039}_{-0.0047}$($2\nu\beta\beta$) $\pm$ 0.029($0\nu\beta\beta$) $\pm$ 0.0098($stab$)\\
DS6     	 		& 0.9014	$\pm$ 0.0011($stat$) $^{+0.0069}_{-0.0081}$($roi$) $^{+0.0023}_{-0.0024}$($2\nu\beta\beta$) $\pm$ 0.029($0\nu\beta\beta$) $\pm$ 0.0090($stab$)\\
\noalign{\smallskip}\hline
\end{tabular}
\caption{\it $AvsE$ cut efficiency and uncertainty contributions for every data-set.}
\label{tab:sys}
\end{center}
\end{table*}

\subsection{Statistical uncertainty}
\label{sec5.1}

The statistical uncertainty of the DEP survival fraction is calculated channel by channel and then averaged, as explained in Section~\ref{sec4}. Results are shown in Table~\ref{tab:mean}. The magnitude changes among data-sets because not all have the same number of events. For instance, the DS2 calibration is shorter than the others while DS6 includes averaging across five long calibrations.

\subsection{Uncertainty due to $AvsE$ energy dependence} 
\label{sec5.2}

The uncertainty from the $AvsE$ energy dependence accounts for the shift in the $AvsE$ distribution between the DEP and the ROI. First, the $AvsE$ mode is calculated for events at the DEP, $\mu_{DEP}$, and events at the ROI, $\mu_{ROI}$. Then, the $AvsE$ cut is varied positively and negatively by the difference of these two values ($\Delta\mu=\mid\mu_{ROI}-\mu_{DEP}\mid$), which is typically $<$0.1 (with the cut value at -1). Finally, the difference in the efficiency is considered as the systematic uncertainty. This systematic is calculated from calibration data channel by channel and then averaged.

\subsection{Uncertainty due to the residual differences between calibration and physics data}
\label{sec5.3}

The uncertainty due to the residual differences between calibration and physics data accounts for the shift in the $AvsE$ distribution between the calibration DEP and the physics data $2\nu\beta\beta$ continuum. The $2\nu\beta\beta$ decay is homogeneously distributed allowing for a cross check of the signal detection efficiency. The $2\nu\beta\beta$ region considered is 950\,keV - 1400 keV to avoid peaks or DCR events. There are not enough background statistics to perform a channel by channel estimation, so the $AvsE$ cut is varied by the difference between the fitted mean $AvsE$ value for calibration DEP events and background events in the $2\nu\beta\beta$ region for all the operating detectors in each data-set ($\Delta\mu=\mid\mu_{2\nu\beta\beta}-\mu_{DEP}\mid$), which is typically $<$0.1. Finally, the difference in the efficiency is considered as the systematic uncertainty. This is the only uncertainty contribution depending on physics data and was updated for the early data-sets after data unblinding. The DS5 data subsets are quoted with the same uncertainty contribution because they reference the same long calibration data.

\subsection{Uncertainty due to the differences between $0\nu\beta\beta$ and DEP events}
\label{sec5.4}

The uncertainty due to the differences between $0\nu\beta\beta$ and DEP events is assessed from a pulse shape simulation. A full waveform simulation was used to validate the performance of the $AvsE$ PSA parameter and estimate the systematic uncertainty due to the difference in $0\nu\beta\beta$ and DEP event populations.  This simulation is based on the standard \textsc{MaGe} \cite{mage} simulation running on \textsc{Geant4.10.3}~\cite{geant4}. We simulate $^{228}$Th-chain calibration events in the calibration track geometry~\cite{calibration} and $0\nu\beta\beta$ events in the enriched detectors.  The  simulation postprocessing framework converts the \textsc{MaGe} output into waveforms using the \textsc{siggen}~\cite{siggen} detector signal simulation and fit waveform shaping parameters~\cite{shanks}.  The simulated waveforms are then processed with the same analysis as the data. The waveform simulation is only available for two detectors from DS1 as the computationally-intensive data waveform fitting has not been expanded to the entire array or all data-sets. Although the fits are based on a single calibration set, comparisons to other data-sets help constrain the variability of the simulation; long calibration runs in DS1, DS3 and DS6 are therefore used as the data reference. The $AvsE$ parameter calibration is performed on the processed simulation data as in the experiment data, but the $j$ cut value is varied between the experimental values from the different data-sets.  

The systematic uncertainty is estimated from two measurements - the agreement between data and simulation at the DEP and the agreement between DEP and $0\nu\beta\beta$ in simulation.  For the first, we assess the difference between each data-set's DEP efficiency and the DEP simulation efficiency at the respective data-set's cut value; these differences range from -1.8\% to +2.3\%.  For the second, we assess the difference between the DEP and $0\nu\beta\beta$ efficiency at each data-set's cut value; these differences range from -0.4\% at the simulation cut value to +1.7\%. Due to the limited number of detector and calibration data-set parings available, the error we estimate is from our most conservative values.  A 2.3\% error for the DEP agreement and 1.7\% for the simulation DEP to $0\nu\beta\beta$ agreement, added in quadrature for a total systematic uncertainty of 2.9\%.  With waveform shaping parameters fit for additional detectors and variation between data-sets taken into account, this systematic error will be better understood and reduced in future analyses.

\subsection{Uncertainty due to time stability}
\label{sec5.5}

The uncertainty due to time stability accounts for variation in the average DEP acceptance observed across all weekly calibration sets. The peak energy window used is 1585.5 keV - 1599.5\,keV, the window with sidebands is 1575 keV - 1610\,keV. For each calibration subset, we compute the efficiency as explained in Section~\ref{sec4}. 

\begin {figure}[ht]
\includegraphics[width=0.45\textwidth]{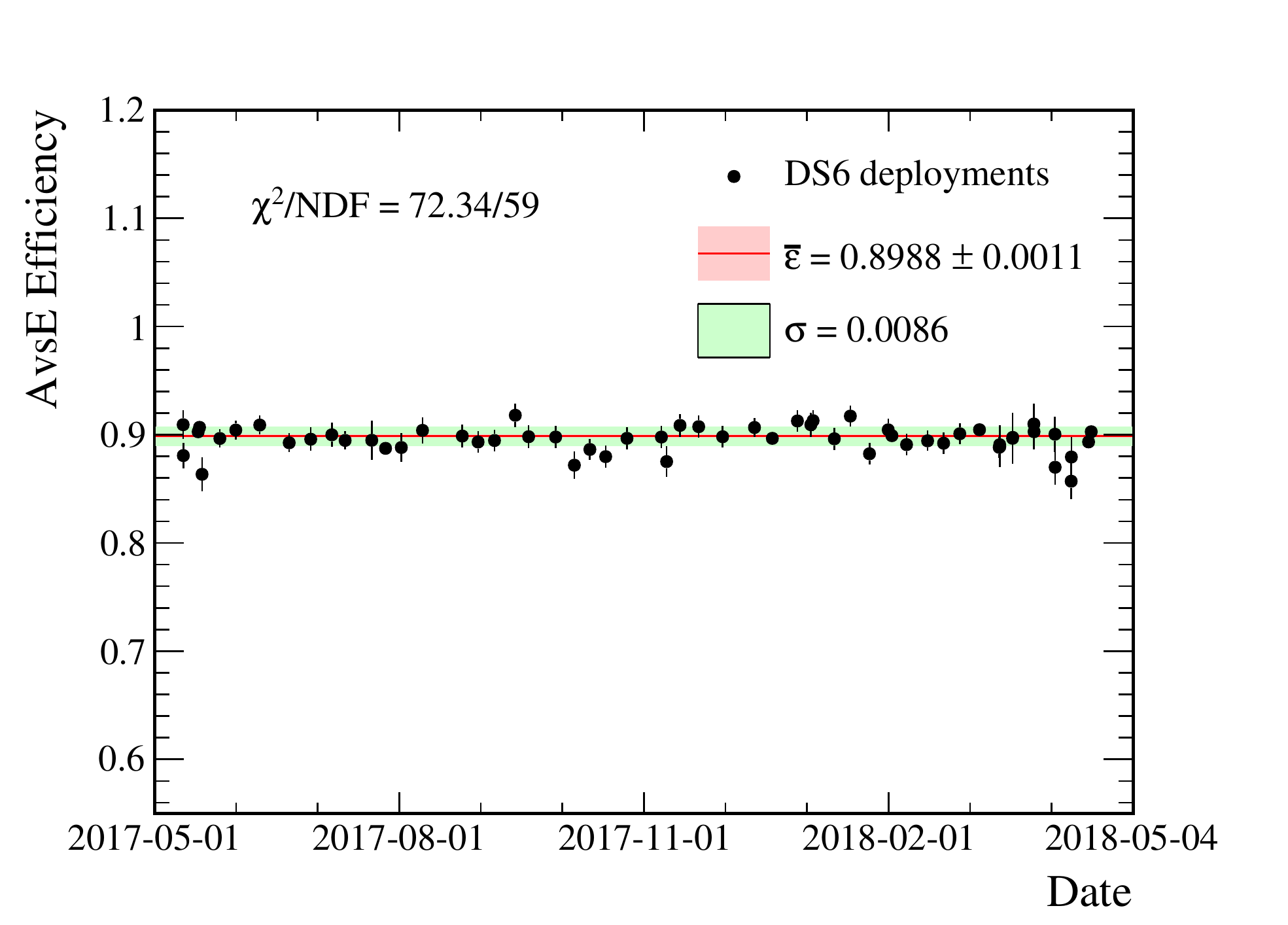}
\centering \caption{\it $AvsE$ stability over the yearlong DS6 dataset. The stability uncertainty uses the weighted standard deviation ($\sigma$), which is significantly larger than the uncertainty on the average.}
\label{fig:stab2}
\end {figure}

The efficiencies over a yearlong period are shown in Fig.~\ref{fig:stab2}. A flat line was fit to the data to compute the weighted average efficiency. As all calibration data are used in this case, the efficiency values differ slightly from those reported in Table~\ref{tab:mean}; this difference ($\Delta\epsilon$) was taken to represent a component of the time stability systematic uncertainty. In all data-sets a non-statistical spread about the weighted average efficiency is observed. To account for such potentially large fluctuations, we conservatively use the weighted standard deviation of the deployment-by-deployment efficiencies ($\sigma$) as the second contribution to the time stability systematic instead of the uncertainty on the weighted average efficiency. Results are given in Table \ref{tab:stab}.

\begin{table}[ht]
\begin{center}
\begin{tabular}{llll}
\hline\noalign{\smallskip}
Data-set	& $\Delta\epsilon (\%)$ & $\sigma (\%)$ & $\sigma_{tot} (\%)$\\
\hline\noalign{\smallskip}
DS0 				& 0.85	& 0.81	& 1.17\\
DS1 				& 0.75	& 1.96	& 2.10\\
DS2 				& 1.13	& 1.32	& 1.87\\
DS3 				& 0.14	& 0.78	& 0.79\\
DS4 				& 0.23	& 0.78	& 0.81\\
DS5a       		& 0.79	& 1.58	& 1.77\\
DS5b       		& 0.44	& 1.10	& 1.18\\
DS5c       		& 0.16	& 0.97	& 0.98\\
DS6       		& 0.26	& 0.86	& 0.90\\
\noalign{\smallskip}\hline
\end{tabular}
\caption{\it Difference between average efficiencies ($\Delta\epsilon$), and scatter of efficiency measurements from all weekly calibrations ($\sigma$), which combine to make the total stability uncertainty ($\sigma_{tot}$).}
\label{tab:stab}
\end{center}
\end{table}

\subsection{Summary of uncertainties}
\label{sec5.6}

The full uncertainties are detailed in Table~\ref{tab:sys} and summarized in Table~\ref{tab:sys2}. Note that DS5 is split into three sub sets (5a, 5b, and 5c) with separate stability systematic uncertainties for consistency with the first result~\cite{mjd0nbb}, but the same cut can be applied in all three cases. Considering all data-sets, the retaining efficiency of single-site events is (90\,$\pm$3.5)\%.

\begin{table}[ht]
\begin{center}
\begin{tabular}{ll}
\hline\noalign{\smallskip}
Data-set	& DEP efficiency\\
\hline\noalign{\smallskip}
DS0 				& 0.9009 $^{+0.0322}_{-0.0353}$\\
DS1 				& 0.9014 $^{+0.0362}_{-0.0405}$\\
DS2 				& 0.9034	$^{+0.0362}_{-0.0385}$\\
DS3 				& 0.8999	$^{+0.0304}_{-0.0314}$\\
DS4 				& 0.8997	$^{+0.0325}_{-0.0367}$\\
DS5a      		& 0.9000	$^{+0.0344}_{-0.0362}$\\
DS5b      		& 0.9000	$^{+0.0318}_{-0.0337}$\\
DS5c      		& 0.9000	$^{+0.0311}_{-0.0331}$\\
DS6      			& 0.9014	$^{+0.0312}_{-0.0315}$\\
\noalign{\smallskip}\hline
\end{tabular}
\caption{\it $AvsE$ cut efficiency (fraction of accepted events from the DEP after background subtraction) and statistical and systematic uncertainty from a quadrature sum of the different contributions.}
\label{tab:sys2}
\end{center}
\end{table}

\section{Background reduction with the $AvsE$ cut}
\label{sec6}

\begin {figure}[ht]
\includegraphics[width=0.45\textwidth]{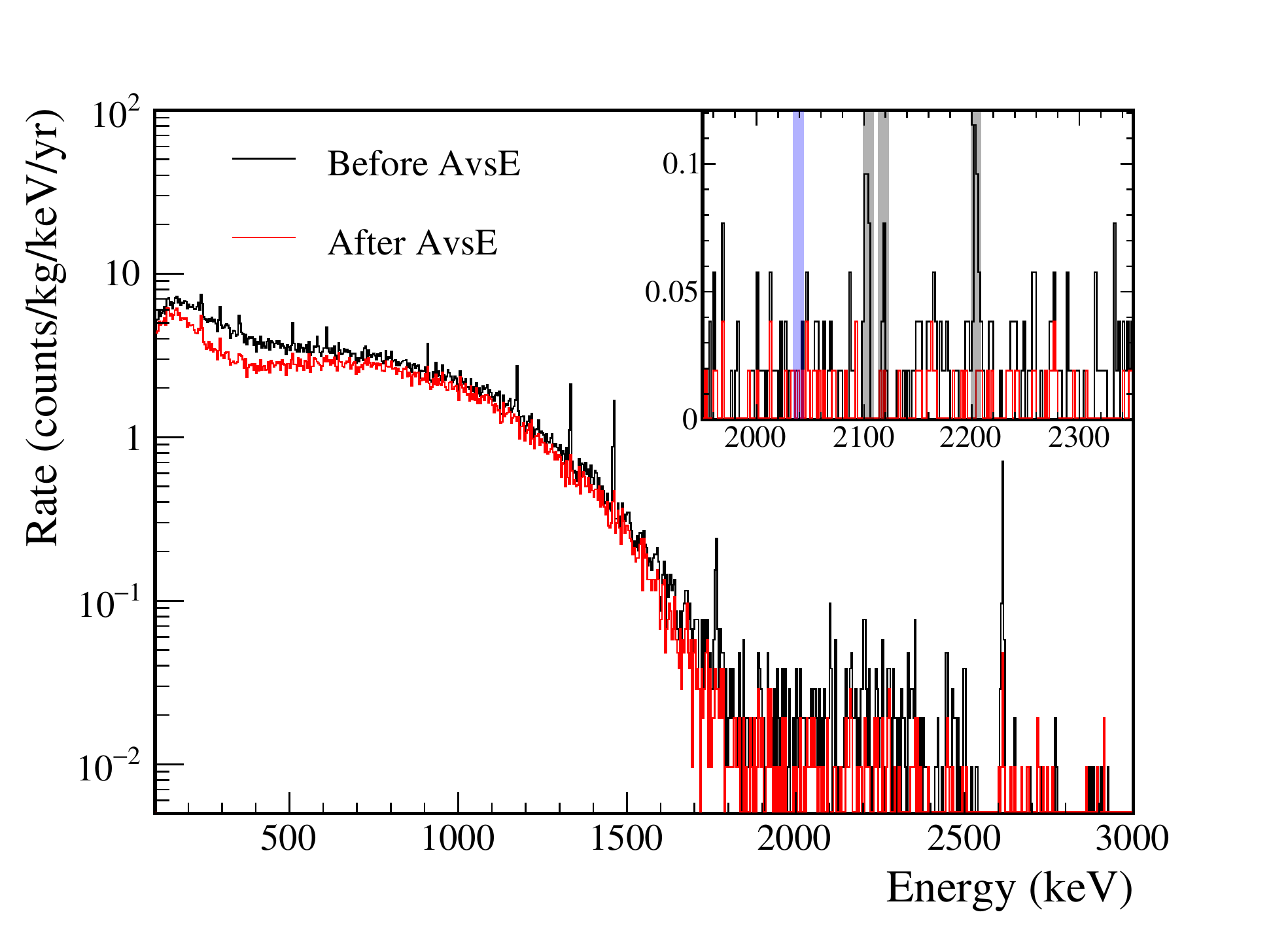}
\centering \caption{\it DS0-6 energy spectrum corresponding to 26\,kg\,yr exposure before and after the $AvsE$ cut.}
\label{fig:cut}
\end {figure}

The $AvsE$ cut is applied to the background data in addition to a standard suite of cuts.  Periods of high noise associated with liquid nitrogen fills or unstable operation are removed.  Non-physical waveforms and pulser events are then removed by data reduction cuts. Multi-detector events caused by multi-site backgrounds across the array are removed by event coincidence with triggers in other germanium detectors or the muon veto.  Surface alpha backgrounds are removed with the DCR cut.  Figure~\ref{fig:cut} shows the effect of the $AvsE$ cut on background data for the full 26~kg\,yr exposure after all these cuts. Table~\ref{tab:bkg} shows the number of events that pass the cut in different energy regions.

\begin{table}[ht]
\centering
\begin{tabular}{lll}
\hline
Energy (keV) & Source & Acceptance \\
\hline
511 & $e^+e^-$, $^{208}$Tl & 0.322\,$\pm$\,0.094 \\
583 & $^{208}$Tl & 0.144\,$\pm$\,0.179 \\
609 & $^{214}$Bi & 0.175\,$\pm$\,0.125 \\
911 & $^{228}$Ac & 0.313\,$\pm$\,0.113 \\
1173 & $^{60}$Co & 0.020\,$\pm$\,0.089 \\
1333 & $^{60}$Co & 0.098\,$\pm$\,0.047 \\
1461 & $^{40}$K & 0.143\,$\pm$\,0.041 \\
1765 & $^{214}$Bi & 0.025\,$\pm$\,0.077 \\
2615 & $^{208}$Tl & 0.062\,$\pm$\,0.028 \\
\hline
1000-1400 & $^{76}$Ge($2\nu\beta\beta$) & 0.860$\pm$0.003 \\
1950-2350* & background window & 0.316$\pm$0.035 \\
\hline
\end{tabular}
\caption{Acceptance of $AvsE$ cut for gamma lines and continuum regions in combined background spectrum.  The acceptance around the gamma lines is calculated with appropriate background subtraction to correct out the continuum contribution.  The background window is a 360~keV window with 10\,keV regions excluded around $Q_{\beta\beta}$ and known gamma lines, as depicted in Fig.~\ref{fig:cut}.}
\label{tab:bkg}
\end{table}

The full energy gamma lines throughout the spectrum are strongly suppressed with the application of the $AvsE$ cut.  The acceptance of the 2$\nu\beta\beta$ spectrum is near 90\%, consistent with expectation for this SSE sample. The known gamma lines (2104, 2118, and 2204\,keV) within the 1950-2350\,keV background averaging window are clearly visible in the initial spectrum, but effectively removed by the $AvsE$ cut; this motivates the removal of these 10\,keV windows from the background window. In the 360\,keV background averaging window (additionally $\pm$5\,keV around $Q_{\beta\beta}$ at 2039\,keV is removed), the $AvsE$ cut provides a factor of 3 suppression of the background index. This reduces the expected background in the optimal window from $\sim$2 to the value obtained of 0.66. Two additional events are present in the $Q_{\beta\beta}$ $\pm$5\,keV window before the cut.

\section*{Conclusions}

The \mj\ Collaboration is operating an array of high purity Ge detectors to search for \BBz\ in \ge.  The PSA implemented to reject multi-site events is known as $AvsE$ and profits from the point contact detector technology. By comparing the maximum amplitude of the current pulse with the energy, events that have a spread-out current pulse and are likely multi-site are rejected by cutting low values of $A$ relative to $E$. This cut is tuned with the DEP of $^{208}$Tl,  whose events have single-site structure like that expected of \BBz. MSE are rejected with $>$90\% efficiency by this cut while SSE are preserved with (90\,$\pm$\,3.5)\% efficiency. The efficiency uncertainty accounts for channel, energy and time-variation, as well as for the position distribution difference between calibration and \BBz\ events, established using simulations.

\section*{Acknowledgments}
This material is based upon work supported by the U.S.~Department of Energy, Office of Science, Office of Nuclear Physics under Award Numbers DE-AC02-05CH11231,  DE-AC05-00OR22725, DE-AC05-76RL0130, DE-AC52-06NA25396, DE-FG02-97ER41020, DE-FG02-97ER41033, DE-FG02-97ER41041, DE-SC0010254, DE-SC0012612, DE-SC0014445, and DE-SC0018060. We acknowledge support from the Particle Astrophysics Program and Nuclear Physics Program of the National Science Foundation through grant numbers MRI-0923142, PHY-1003399, PHY-1102292, PHY-1206314, PHY-1614611, PHY-1812409, and PHY-1812356. We gratefully acknowledge the support of the U.S.~Department of Energy through the LANL/LDRD Program and through the PNNL/LDRD Program for this work. We acknowledge support from the Russian Foundation for Basic Research, grant No.~15-02-02919. We acknowledge the support of the Natural Sciences and Engineering Research Council of Canada, funding reference number SAPIN-2017-00023, and from the Canada Foundation from Innovation John R.~Evans Leaders Fund.  This research used resources provided by the Oak Ridge Leadership Computing Facility at Oak Ridge National Laboratory and by the National Energy Research Scientific Computing Center, a U.S.~Department of Energy Office of Science User Facility. We thank our hosts and colleagues at the Sanford Underground Research Facility for their support.

\bibliography{avse}

\end{document}